\documentclass[reprint,superscriptaddress]{revtex4-1}
\usepackage{amsmath}
\usepackage{graphicx}
\usepackage{float}
\usepackage{cancel}
\usepackage[section]{placeins}

\begin{document}

\title{Ab-initio calculations of the linear and nonlinear
  susceptibilities of N$_2$, O$_2$, and air in the mid-infrared}

\author{Jeffrey M. Brown}
\affiliation{Department of Physics and Astronomy, Louisiana State
  University, Baton Rouge, LA 70803-4001, USA}
\affiliation{Centre de Physique Th\'eorique, Ecole Polytechnique, CNRS, Universit\'e
  Paris-Saclay, Route de Saclay, F-91128 Palaiseau, France}

\author{Arnaud Couairon}
\affiliation{Centre de Physique Th\'eorique, Ecole Polytechnique, CNRS, Universit\'e
  Paris-Saclay, Route de Saclay, F-91128 Palaiseau, France}

\author{Mette B. Gaarde}
\affiliation{Department of Physics and Astronomy, Louisiana State
  University, Baton Rouge, LA 70803-4001, USA}

\begin{abstract}
  We present first-principles calculations of the linear and nonlinear
  susceptibilities of N$_2$, O$_2$, and air in the mid-infrared
  wavelength regime, from $1-4$ $\mu$m. We extract the
  frequency-dependent susceptibilities from the full time-dependent
  dipole moment that is calculated using time-dependent density
  functional theory. We find good agreement with curves derived from
  experimental results for the linear susceptibility, and with
  measurements for the nonlinear susceptibility up to 2.4 $\mu$m. We
  also find that the susceptibilities are insensitive to the laser
  intensity even in the strong field regime up to $5\times 10^{13}$
  W/cm$^2$. Our results will allow accurate calculations of the
  long-distance propagation of intense MIR laser pulses in air.
\end{abstract}

\maketitle

\section{Motivation}
\label{sec:motivation}

The propagation of ultrashort, intense laser pulses in gaseous media
has been extensively studied in the visible and near infrared
wavelength regions \cite{couairon-filamentation}.  The nonlinear
optical phenomena associated with these pulses arise from a
combination of the medium's linear optical properties (dispersion) and
intensity-dependent nonlinear processes, such as Kerr self-focusing
and ionization.  With the new availability of ultrashort pulse laser
sources in the mid-infrared (MIR, $2-4$ $\mu$m) we enter a new frontier
in ultrafast science, where many applications in strong field physics
benefit greatly from an increase of the quiver energy of the electron
in the longer wavelength laser field \cite{agostini:2008}. However,
there are still many open questions regarding the roles of various
nonlinear processes that drive the long range propagation of MIR
pulses. It has been proposed that the relative influence of
dispersion, self-focusing, and ionization may be different than those
for near-infrared wavelengths \cite{panagiotopoulos2015super}.
Therefore, an accurate description of the linear and nonlinear optical
properties of common gases in the atmosphere is crucial for predictive
modeling of the long-range propagation of MIR laser pulses.

The linear optical properties of molecular nitrogen N$_2$ and
molecular oxygen O$_2$ in the visible spectrum have been known since
the 1960's \cite{Peck:66, peck1972dispersion, mizrahi1985dispersion}
and have typically been modeled empirically in the visible region
using Sellmeier-like equations. Newer measurements \cite{Ciddor:96,
  Zhang:08, Kren:11} have helped improve and extend the modeling up to
2 $\mu$m. The nonlinear optical properties of these species have also
been measured \cite{shelton1994measurements, zahedpour2015n2} up to
2.4 $\mu$m. Of particular interest is the value of the nonlinear index
coefficient $n_2$, where the total index of the medium $n$ has a
dependence on the instantaneous intensity of the laser $I$ through
relation $n = n_0 + n_2 I$.  Experiments
\cite{mitrofanov2015mid,panov2016supercontinuum} and simulations
\cite{panagiotopoulos2015super} utilizing MIR laser wavelengths call
for new investigations on the linear and nonlinear properties of the
constituents of air above 2.4 $\mu$m.

Multiple theoretical approaches have been proposed for determining
these optical properties. A Kramers-Kr\"onig transformation of the
multiphoton absorption rate led to the prediction of the dispersion of
$n_2$ for noble gases in the mid-infrared
\cite{bree2010method,bree2012kramers}. Ab-initio multiconfiguration
self-consistent field (MCSCF) cubic response theory calculations were
performed to extract hyperpolarizability and subsequently the
frequency dependence of $n_2$ for multi-ionized noble gases
\cite{tarazkar2016nonlinear} and for N$_2$
\cite{tarazkar2015theoretical}. Calculations of the nonlinear response
of O$_2$ in the mid-infrared seem to be relatively unexplored.

In this paper we calculate the linear and nonlinear optical properties
of N$_2$ and O$_2$ molecules for wavelengths ranging from $1-4$
$\mu$m. This is done using time-dependent density functional theory
(TDDFT), as implemented in the software package Octopus
\cite{octopus,octopus-parallel}, to calculate the multi-electron
dipole response to a short, intense laser pulse. From the resulting
dipole spectrum, it is possible to extract the linear and nonlinear
optical properties of both gas species. We find that the extracted
values for the linear index $n_0$ and nonlinear index $n_2$ of both
species are independent of the laser intensities with the range
$10^{10}-5\times 10^{13}$ W/cm$^2$ and that the values are in good
agreement with published experimental data between $1-2.4$ $\mu$m. We
infer the linear and nonlinear optical properties of air from
corresponding calculations for its constituents.

The outline of the paper is as follows. Section \ref{sec:simulations}
details the calculation of the time-dependent dipole moment using
TDDFT. Section \ref{sec:suscept} describes how the macroscopic linear
and nonlinear susceptibilities are extracted from the microscopic
time-dependent dipole moment. In Section \ref{sec:results} the results
of the calculated linear and nonlinear refractive indices are
presented and compared to available experimental data, followed by a
summary in Section \ref{sec:summary}.

\section{Simulations}
\label{sec:simulations}
We simulate the multi-electron dynamics of N$_2$ and O$_2$ using TDDFT
as implemented in the open source software package Octopus
\cite{octopus}. Non-relativistic Kohn-Sham density functional theory
allows an interacting many-electron system to be represented by an
auxiliary system of non-interacting electron densities where both
systems have the same ground state charge density.  The Hamiltonian of
the non-interacting system is written as the sum of the kinetic energy
operator $T$ and the Kohn-Sham potential $V_{KS}$:
$H = T + V_{KS}[\rho(r,t)]$. The Kohn-Sham potential is a functional
of the electron density $\rho$ that is separated into
$V_{KS}[\rho]=V_{ext} + V_{H}[\rho] + V_{XC}[\rho]$, where $V_{ext}$
is the external potential, $V_{H}$ is the Hartree potential
representing electrostatic interaction between electrons, and $V_{XC}$
is the exchange-correlation operator that contains all non-trivial
interactions. The exact form of $V_{XC}$ is unknown and is therefore
approximated to various levels of sophistication. For time-dependent
calculations of the molecules interacting with the laser field, the
adiabatic approximation is made and assumes that the
exchange-correlation potential is time independent.

The simulations take place in two steps. The first is to determine the
ground state through minimizing the total energy of the
system. Convergence of the ground state energy to obtain a realistic
value of the ionization potential $I_p$ is important, since the
energies of the high-lying occupied molecular orbitals determine much
of the optical properties of the molecule. Once a suitable ground
state has been found, the second step is a time-dependent calculation
of the dipole moment of the total electronic response of the molecule
as it interacts with the MIR laser pulse.

To achieve an accurate convergence to the ground state, each molecule
requires a different set of simulation parameters. The only common
parameters between the two molecular simulations are that the default
pseudo-potentials provided with Octopus are used and that both
molecules live on a cylindrical grid with dimensions length = 30,
radius = 15, and a grid spacing = 0.3 (atomic units are used
throughout unless otherwise specified). The large length is necessary
to avoid boundary effects since long wavelength pulses can accelerate
the electrons far from the origin during the time-dependent portion of
the simulation \cite{agostini:2008,krause}. Since only the lower order
harmonic response is needed for calculating the first- and third-order
susceptibilities of the medium, the simulation parameters are chosen
such that the dipole spectrum is converged up to and including
harmonic 7.

For N$_2$ it is sufficient to run Octopus simulations in
spin-unpolarized mode, which places two electrons in each
orbital. This effectively forces the same energy on both spin-up and
spin-down electrons, reducing the computational cost by half. For
N$_2$, a bond length of 2.068 was found to minimize the total energy
of the system. The exchange-correlation (XC) functionals in the local
density approximation (LDA) \cite{dirac_1930,Bloch1929,correlation-pz}
work quite well with the addition of the self-interaction correction
ADSIC \cite{adsic}. From this configuration, the ground state orbital
energies match closely to the experimentally measured ones
\cite{nitrogen-orbital-energies} (Table \ref{tab:nitrogen-energies}).

\begin{table}[h]
  \begin{ruledtabular}
    \begin{tabular}{llll}
    MO          & Occ & Exp     & Sim\\
    \hline
    2$\sigma_g$ & 2   & 1.533   & 1.299\\
    2$\sigma_u$ & 2   & 0.7717  & 0.7029\\
    1$\pi_u$    & 4   & 0.6273  & 0.6760\\
    3$\sigma_g$ & 2   & 0.5726  & 0.5953
    \end{tabular}
    \end{ruledtabular}
    \caption{\label{tab:nitrogen-energies}N$_2$ molecular orbitals (MO),
    occupation numbers (Occ), and energies for experimental (Exp) and
    calculated (Sim) values (atomic units).}
\end{table}


The ground state of O$_2$, commonly known as triplet oxygen, contains
two unpaired, spin-up electrons occupying two $\pi_g$ molecular
orbitals. Therefore, it is necessary to run Octopus in spin-polarized
mode, where spin-up and spin-down electrons are placed in their own
orbitals and allowed to evolve independently in energy.

For O$_2$, a bond length of 2.2866 was found to minimize the energy of
the system.  Using the GGA exchange-correlation functionals
(\texttt{XCFunctional = gga\_x\_lb + gga\_c\_tca})
\cite{van1994exchange, tognetti2008new}, we find good agreement
between the calculated and measured orbital energies
\cite{oxygen-orbital-energies} (Table \ref{tab:oxygen-energies}). A
number of other exchange-correlation functionals with varying levels
of complexity were explored, including LDA and hybrid functionals
([box]3lyp, PBE0, M05), but none of these other options produced a
ground state with an ionization potential $I_p$ within 15\% of the
measured value.


\begin{table}[h]
  \begin{ruledtabular}
  \begin{tabular}{llll}
    MO           & Occ & Exp     & Sim (up, dn) \\
    \hline
    2$\sigma_g$  & 2   & 1.697   & 1.452, 1.387 \\
    2$\sigma_u $ & 2   & 1.096   & 0.9445, 0.8770 \\
    1$\pi_u$     & 4   & 0.7218  & 0.7323, 0.6664 \\
    3$\sigma_g$  & 2   & 0.7273  & 0.7252, 0.6658 \\
    1$\pi_g$     & 2   & 0.4436  & 0.4688, 0.3966
  \end{tabular}
  \end{ruledtabular}
  \caption{O$_2$ molecular orbitals (MO), occupation numbers (Occ),
    and energies for experimental (Exp) and calculated (Sim, spin-up
    and spin-down) values (atomic units).}
  \label{tab:oxygen-energies}
\end{table}

For the time-dependent calculation, we calculate the response to a
few-cycle, linearly polarized, MIR laser pulse given by
\begin{equation}
  \label{eq:field}
  E(t) = E_0 \sin^{10}(\omega t / 2 N_c) \sin(\omega t)\ ,
\end{equation}
where the field strength $E_0 = \sqrt{2I_0 / \epsilon_0 c}$ varies
through the peak intensities $I_0 = 10^{10}-10^{14}$ W/cm$^2$, $N_c=8$
is the number of half cycles under the envelope, and
$\omega = 2\pi c / \lambda$ with wavelengths $\lambda$ corresponding
to $1-4$ $\mu$m.  In order to minimize artifacts from portions of the
electron density nearing the edges of the computational box, complex
absorbing boundary conditions are added.

\section{Calculation of susceptibilities}
\label{sec:suscept}
\begin{figure}[h]
  \centering
  \includegraphics[width=0.8\linewidth]{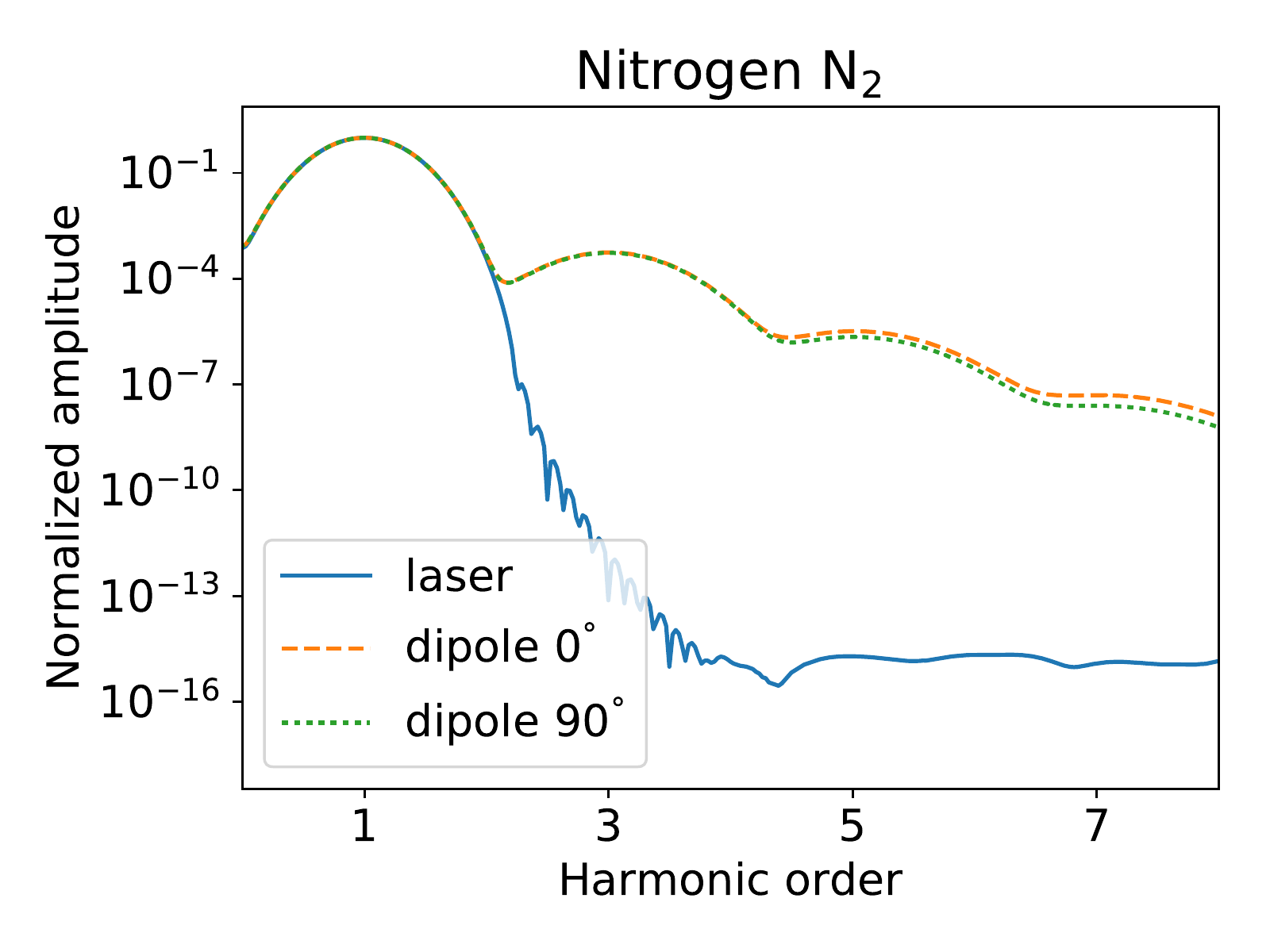}
  \caption{The spectral amplitude of the laser pulse and the induced
    dipole moment of an N$_2$ molecule for two molecular
    orientations. 0$^\circ$ indicates that the polarization of the
    laser is aligned along the molecular axis, while for 90$^\circ$
    they are perpendicular.}
  \label{fig:spectrum}
\end{figure}

The goal of the time-dependent calculations is to extract the
susceptibility of a bulk gaseous medium containing an ensemble of
randomly-oriented molecules. Therefore time-dependent simulations are
performed for many angles
$\theta = [0^\circ, 15^\circ, 30^\circ, \dots, 90^\circ]$ between the
molecular axis and laser polarization.  The dipole spectra are
calculated using the Fourier transform of the time-dependent dipole
moments (Figure \ref{fig:spectrum}).

The resulting magnitude of the dipole spectra at the fundamental laser
frequency is found to have a $\cos^2(\theta)$ dependence even at a
peak laser intensity of $I_0=10^{14}$ W/cm$^2$. Given the
$\cos^2(\theta)$ dependence, it is possible to compute the polarization
spectrum of an ensemble of randomly oriented molecules using a linear
combination of the dipole spectra for parallel and perpendicular
orientations of the molecules:
\begin{equation}
  \label{eq:dipole-para-perp}
  \hat{P}(\omega) = \rho\left(\frac13 \hat{d}_{\parallel}(\omega) + \frac23 \hat{d}_{\perp}(\omega)\right)\ ,
\end{equation}
where $\rho$ is the neutral density of molecules. For N$_2$,
$\rho = 2.688\times 10^{25}$ m$^{-3}$ and for O$_2$
$\rho=2.505\times 10^{25}$ m$^{-3}$ at atmospheric pressure and
room temperature. We note that this linear combination is typically
valid only in the limit of low intensity laser pulses
\cite{Boyd}. However, due to the low amount of ionization (a ground
state population reduction of $\approx 10^{-5}$) that occurs for
mid-infrared pulses even up to $10^{14}$ W/cm$^2$, we find that
Eq. \eqref{eq:dipole-para-perp} is a good approximation.

The total susceptibility of the media, which includes both linear and
nonlinear components, can be calculated by dividing the full
polarization response spectrum $\hat{P}(\omega)$ by the laser's
spectrum $E(\omega)$:
\begin{equation}
  \label{eq:suscept}
  \chi(\omega) = \frac{\hat{P}(\omega)}{\epsilon_0 \hat{E}(\omega)}\ .
\end{equation}
To extract the components of $\chi(\omega)$ corresponding to the
linear $\chi^{(1)}(\omega)$ and nonlinear $\chi^{(3)}(\omega)$
properties of the medium, we employ a procedure that separates the
linear and nonlinear responses spectrally.

In laser pulse propagation simulations the time-dependent polarization
$P(t)$ of the medium is calculated as a power series expansion of odd
harmonics of the field $E(t)$. For example, considering up to 5th
order nonlinear processes, the polarization is
\begin{equation}
  \label{eq:power-series}
  P(t) = \epsilon_0 \left[ \chi^{(1)} E(t) + \chi^{(3)} E^3(t)  + \chi^{(5)} E^5(t) \right]\ .
\end{equation}
We relate the expansion Eq. \eqref{eq:power-series} to the various
harmonics in the dipole spectrum using the Fourier transform, yielding
a set of equations where each polarization spectra $\hat{P}(n\omega)$
can be written as a sum of linear and nonlinear contributions up to
order $\chi^{(5)}$:
\begin{align*}
  \frac{1}{\epsilon_0}&\left[\hat{P}(\omega) + \hat{P}(3\omega) + \hat{P}(5\omega)\right]\\
                      &= \chi^{(1)} \hat{E}_1(\omega) + \chi^{(3)} \hat{E}_3(\omega) + \chi^{(5)} \hat{E}_5(\omega)\\
                      &+ \cancel{\chi^{(1)} \hat{E}_1(3\omega)} + \chi^{(3)} \hat{E}_3(3\omega) + \chi^{(5)} \hat{E}_5(3\omega)\\
                      &+ \cancel{\chi^{(1)} \hat{E}_1(5\omega)} + \cancel{\chi^{(3)} \hat{E}_3(5\omega)} + \chi^{(5)} \hat{E}_5(5\omega)\ ,
\end{align*}
where the quantities $\hat{E}_n(\omega)$ represent the Fourier
transform of powers of the field $E^n(t)$. Terms containing no signal
at a particular frequency are set to zero; for example there is no 3rd
harmonic in the fundamental field and therefore $E_1(3\omega) =
0$. Collecting terms of the same harmonic order yields a set of
equations where each order susceptibility can be written in terms of
the calculated molecular polarizations $P$ and field spectra $E$:
\begin{align}
    \chi^{(5)} &= \frac{\hat{P}(5\omega)}{\epsilon_0 \hat{E}_5(5\omega)}  \label{eq:sus5}\\
    \chi^{(3)} &= \frac{\hat{P}(3\omega)}{\epsilon_0 \hat{E}_3(3\omega)} - \chi^{(5)}\frac{\hat{E}_5(3\omega)}{\hat{E}_3(3\omega)} \label{eq:sus3}\\
  \chi^{(1)} &= \frac{\hat{P}(\omega)}{\epsilon_0 \hat{E}_1(\omega)} - \chi^{(3)}\frac{\hat{E}_3(\omega)}{\hat{E}_1(\omega)} - \chi^{(5)}\frac{\hat{E}_5(\omega)}{\hat{E}_1(\omega)}
               \label{eq:sus1}
\end{align}
Conceptually, this corresponds to, for example, eliminating the
contribution to the third harmonic yield from the fifth order process
that involves absorbing four laser photons and emitting one, etc.  The
procedure is general and does not depend on the particular shape of
the field $E(t)$. It also avoids division by small values since the
polarization at each harmonic order is divided by a spectral field
component that also contains a signal at that particular harmonic.
However, we note that this perturbative approach is limited to
intensity and wavelength regimes where ionization is small.

In practice, it is only necessary to consider nonlinear processes up
to $\chi^{(5)}$ in order to extract intensity-independent values for
$\chi^{(1)}$ and $\chi^{(3)}$. The magnitude of the nonlinear
contribution to the fundamental and third harmonic polarizations drops
off quite rapidly as the harmonic order increases and becomes
negligible for harmonic orders 7 and above.  We obtain
intensity-independent susceptibilities over the wavelength range of
$2-4$ $\mu$m for peak intensities up to $10^{14}$ W/cm$^2$. For
wavelengths between $1-2$ $\mu$m, the extracted $\chi^{(3)}$ begins to
show a small intensity dependence for peak intensity values above
$5\times 10^{13}$ W/cm$^2$ due to a non-negligible amount of
ionization. Intensity limitations of a perturbative approach of
modeling the total susceptibility of a medium has also been observed
in ab-initio calculations of atomic hydrogen
\cite{spott2014perturbativesuscept}.

Using the expressions for susceptibility in
Eqs. \eqref{eq:sus5}-\eqref{eq:sus1}, the linear refractive index is
\begin{equation}
  \label{eq:n0}
  n_0(\omega) = \sqrt{1 + \alpha\chi^{(1)}(\omega)} \ ,
\end{equation}
where $\alpha$ is a scaling factor described in more detail below.
The nonlinear refractive index is
\begin{equation}
  \label{eq:n2}
  n_2(\omega) = \frac34\frac{\chi^{(3)}(\omega)}{\epsilon_0 c n_0^2(\omega)} \ .
\end{equation}

The scaling factor $\alpha$ ($\alpha = 1.055$ for N$_2$ and
$\alpha = 1.034$ for O$_2$) is included to facilitate graphical
comparison between the calculated values of this work and experimental
values.
The percentage adjustment of the linear susceptibility is consistent
with the percentage difference between the calculated and measured
values of $I_p$ for both species,
$\Delta I_p^{\{\mathrm{N}_2\}} = 3.9\%$ and
$\Delta I_p^{\{\mathrm{O}_2\}} = 5.7\%$, where the simulations have
overestimated the binding energy of the highest energy electrons,
resulting in a weaker response to the laser field.

\section{Results}
\label{sec:results}
\begin{figure}[h]
  \centering
  \includegraphics[width=0.8\linewidth]{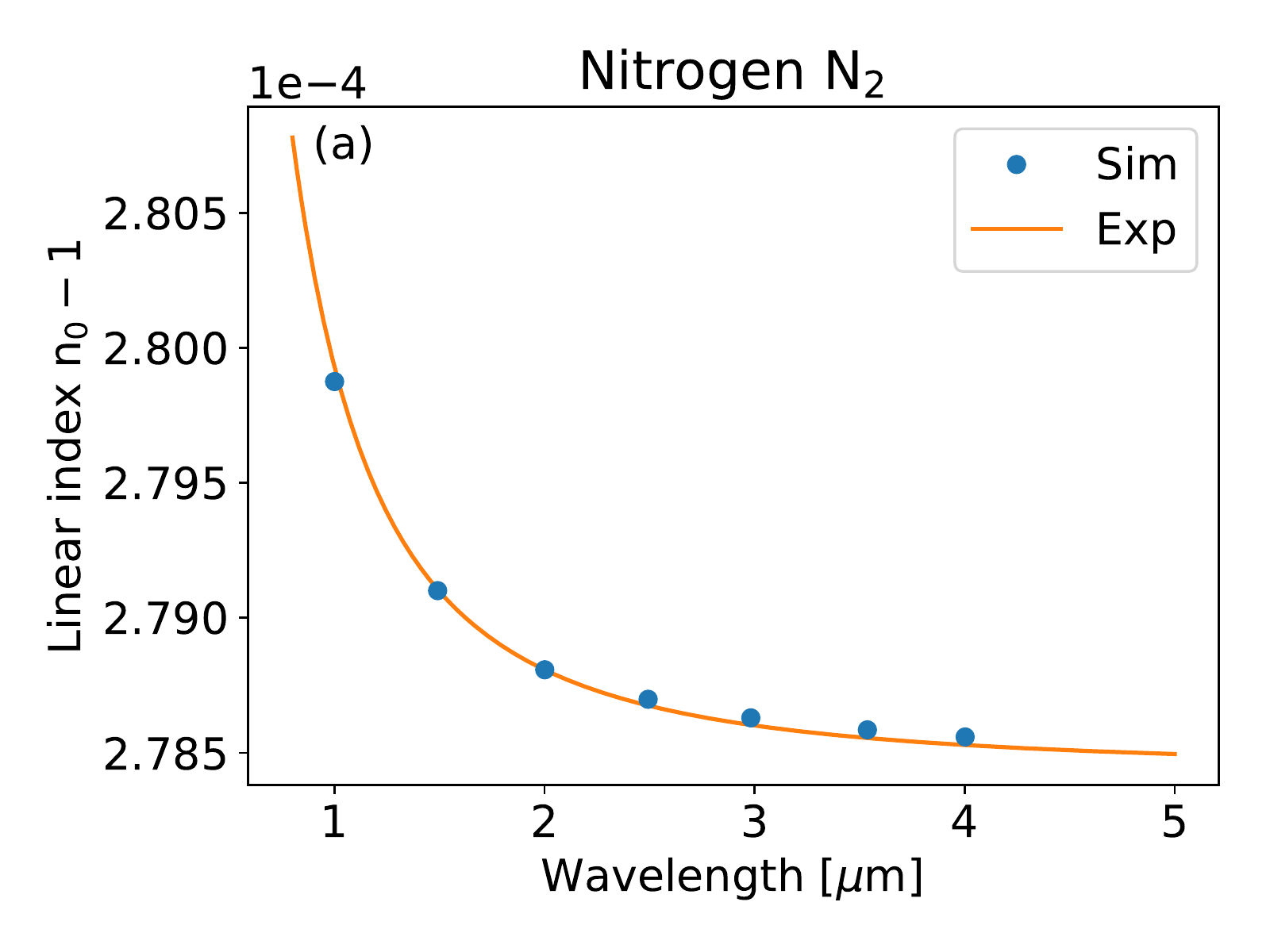}
  \includegraphics[width=0.8\linewidth]{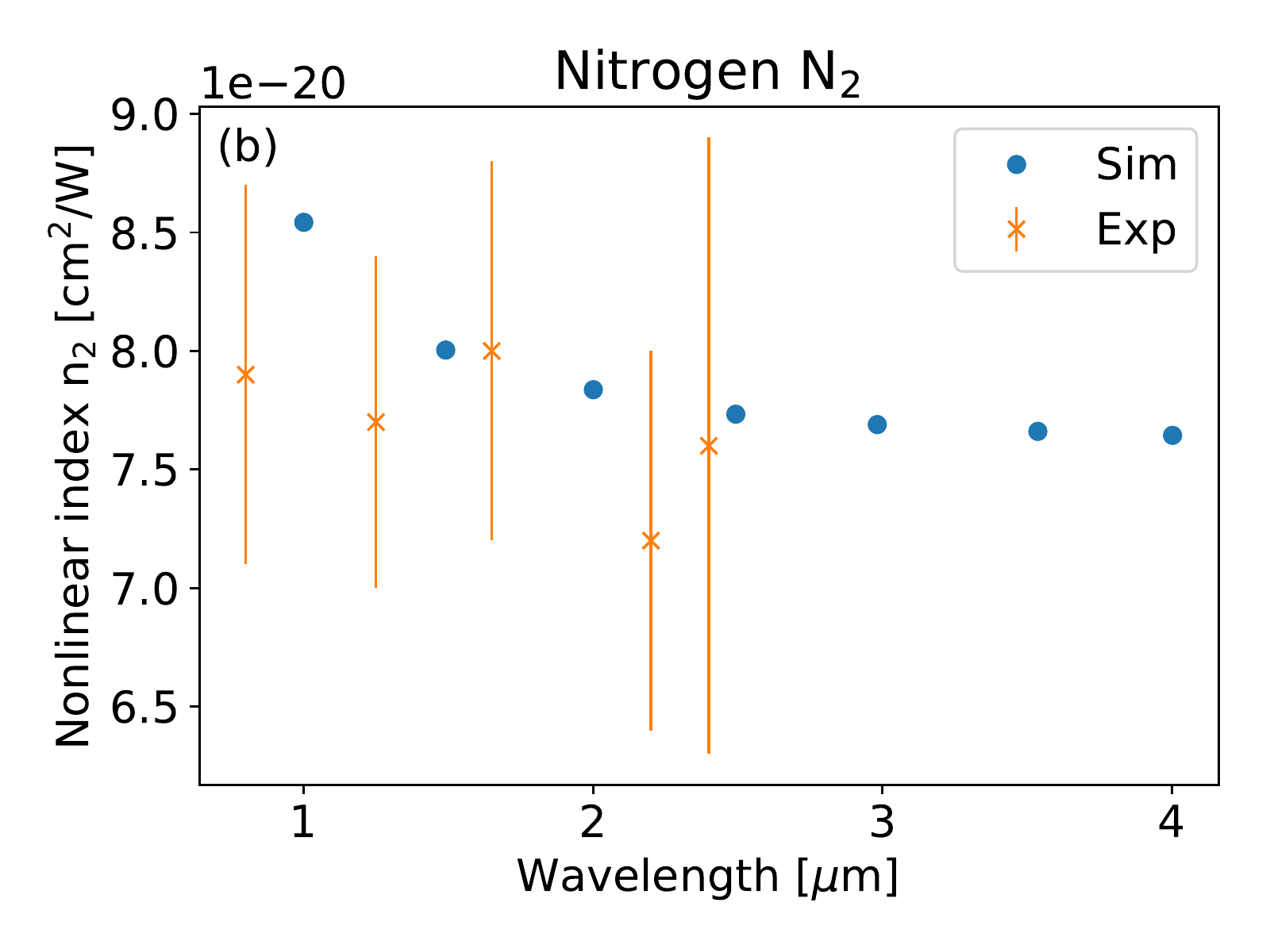}
  \caption{The calculated (Sim) optical properties of N$_2$ compared to
    experimental values (Exp). (a) The calculated index of refraction $n_0$
    is compared to the curve Eq. \eqref{eq:nitrogen-exp}. (b) The
    nonlinear refractive index $n_2$ compared to experimental data
    \cite{zahedpour2015n2}.}
  \label{fig:nitrogen-n0-n2}
\end{figure}

In Figure \ref{fig:nitrogen-n0-n2}a, the calculated values of the
linear index $n_0$ for N$_2$ are compared to the curve derived from
experimental data (Peck and Khanna 1966 \cite{Peck:66})
\begin{equation}
  n(\lambda) = 1 + 6.497378\times 10^{-5} + \frac{3.0738649\times 10^{-2}}{\lambda_0^{-2} - \lambda^{-2}}\ ,
  \label{eq:nitrogen-exp}
\end{equation}
where $\lambda_0^{\{n_0,\mathrm{N}_2\}} = 0.0833$ $\mu$m and is valid from 0.4679 to 2.0586
$\mu$m.  Despite the stated upper limit of 2 $\mu$m,
Eq. \eqref{eq:nitrogen-exp} fits the calculated values remarkably well
up to 4 $\mu$m. In Figure \ref{fig:nitrogen-n0-n2}b, the nonlinear
index is compared to experimental data (Zahedpour 2015
\cite{zahedpour2015n2}) and is also found to be in very close
agreement.

\begin{figure}[h]
  \centering
  \includegraphics[width=0.8\linewidth]{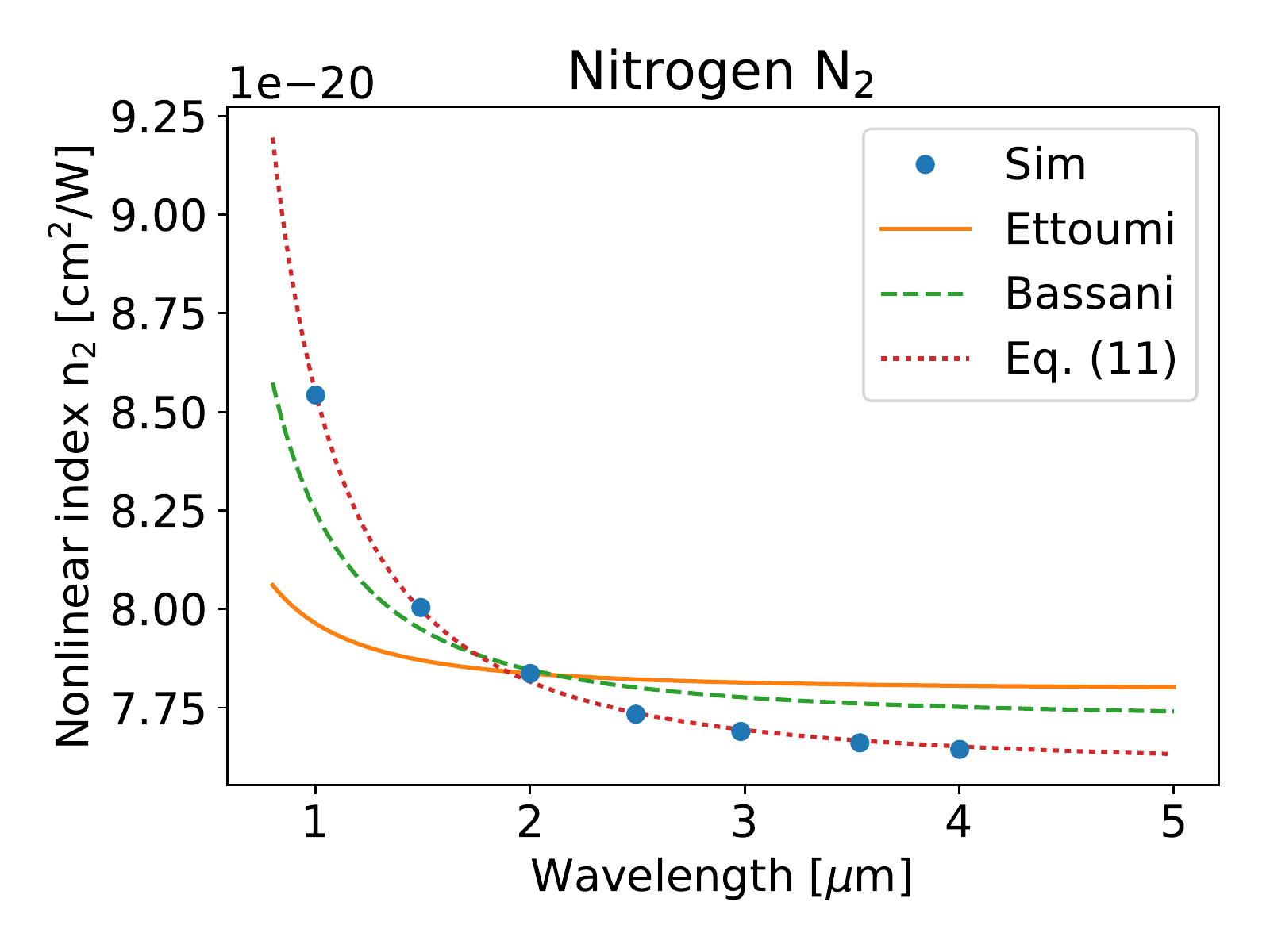}
  \caption{The calculated (Sim) nonlinear refractive index values
    $n_2$ of N$_2$ are compared to two formulations of a generalized
    Miller's rule for third order susceptibilities. The best match for
    these values is obtained by a new Sellmeier-like equation
    Eq. \eqref{eq:new}.}
  \label{fig:nitrogen-miller}
\end{figure}

Since there is a decreasing spectral trend of the calculated values of
$n_2$, it is interesting to compare its curve to the prediction
provided by a generalized Miller's rule for third order
susceptibilities. Two formulations have been proposed in the
literature. The first is
$\chi^{(3)}(\omega) = \chi^{(3)}(\omega_0) \left[\chi^{(1)}(\omega) /
  \chi^{(1)}(\omega_0)\right]^4$ proposed by Ettoumi {\it et al.}
\cite{Ettoumi:10} where $\omega_0$ corresponds to a reference value
(e.g. 2 $\mu$m). The second is
$\chi^{(3)}(\omega) = \delta \chi^{(1)}(3\omega) \chi^{(1)}(\omega)^3$
proposed by Bassani {\it et al.} \cite{bassani1998general} where the
factor $\delta = 2.841\times 10^{-13}$ m$^2$/V$^2$ is determined by performing a
least-squares fit of the simulation data. In Figure
\ref{fig:nitrogen-miller}, these predictive curves of $n_2$ are
plotted along with the values calculated in this work.

It is clear that both predicted curves underestimate the dispersion of
$n_2$ at long wavelengths compared to our calculated values, since
both are ``flatter'' at long wavelengths.  This finding is consistent
with that of Ref. \cite{mizrahi1985dispersion} which pointed out that
Miller's rule tends to underestimate the strength of the dispersion,
and that there is not in general a strong correlation between the
linear and nonlinear dispersion properties over a wide range of gases.

We find that the calculated values of $n_2$ are well fitted by a
Sellmeier-like equation
\begin{equation}
  \label{eq:new}
  n_2(\lambda) = \frac{P^{-1}}{\lambda_0^{-2} - \lambda^{-2}}\ ,
\end{equation}
where $P^{\{\mathrm{N}_2\}} = 14.63$ GW and $\lambda_0^{\{n_2,N_2\}} = 0.3334$ $\mu$m. As
seen in Figure \ref{fig:nitrogen-miller}, Eq. \eqref{eq:new} captures
the dispersion of $n_2$ well, though it does force a singularity at
$\lambda = 0.333$ $\mu$m.

\begin{figure}[h]
  \centering
  \includegraphics[width=0.8\linewidth]{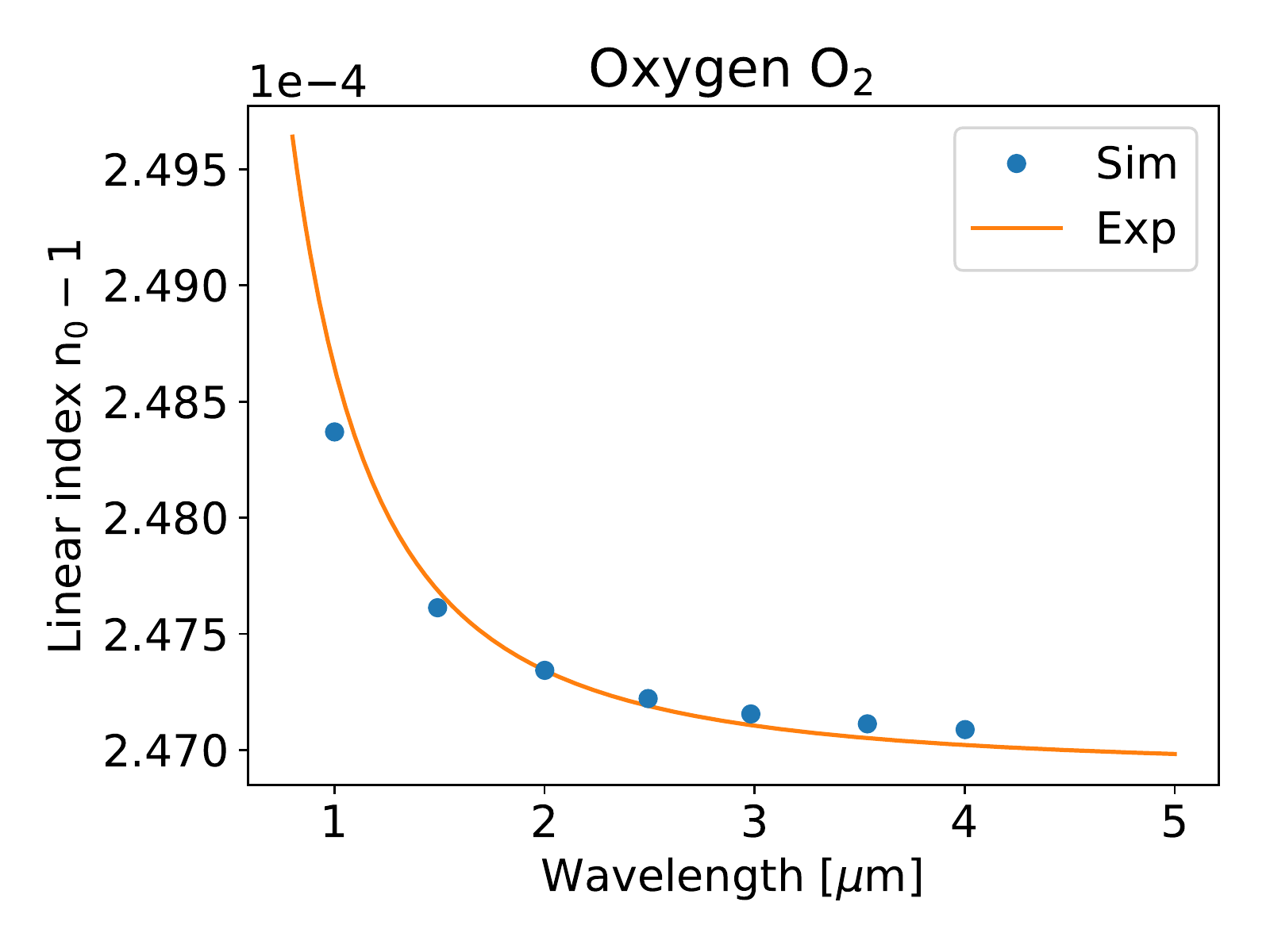}
  \caption{The calculated values (Sim) of the index of refraction
    $n_0$ of O$_2$ are compared to the experimentally derived curve
    Eq. \eqref{eq:oxygen-exp} (Exp).}
  \label{fig:oxygen-n0}
\end{figure}

In Figure \ref{fig:oxygen-n0}, the calculated values for the linear
index $n_0$ for O$_2$ are compared to the experimentally derived curve
(Zhang 2008 \cite{Zhang:08}, Kren 2011 \cite{Kren:11})
\begin{equation}
  n(\lambda) = 1 + 1.181494\times 10^{-4} + \frac{9.708931\times 10^{-3}}{\lambda_0^{-2} - \lambda^{-2}}\ ,
  \label{eq:oxygen-exp}
\end{equation}
where $\lambda_0^{\{n_0,\mathrm{O}_2\}} = 0.115$ $\mu$m and is valid from 0.4 to 1.8
$\mu$m. Extending this curve into the mid-infrared wavelength region
shows that the scaled values of linear index from the simulation are
well represented by Eq. \eqref{eq:oxygen-exp}.
The calculated nonlinear index from the simulations are also in
reasonable agreement with experimental data
\cite{zahedpour2015n2}. However, we do not find an increase of $n_2$
near 2.4 $\mu$m which places our calculated values in closer agreement
with the experimental data from Shelton and Rice \cite{shelton} for
this particular wavelength. Just as with N$_2$, the values of $n_2$
for O$_2$ can be fitted with the Sellmeier-like equation
Eq. \eqref{eq:new} using the parameters $P^{\{\mathrm{O}_2\}} = 14.62$ GW and
$\lambda_0^{\{n_2,\mathrm{O}_2\}} = 0.3360$ $\mu$m.

We note that there is only a few percent difference between the
calculated values of the nonlinear index n$_2$ for N$_2$ and O$_2$ and
that this is merely a coincidence. In general, the value of $I_p$ for
a particular species is not necessarily correlated with its value of
$n_2$.  A well-known example of this is the case of Ar and N$_2$ which
have very similar values of $I_p$, yet $n_2$ for Ar is roughly 25\%
larger than that of N$_2$ in the MIR regime \cite{zahedpour2015n2}.


\begin{figure}[h]
  \centering
  \includegraphics[width=0.8\linewidth]{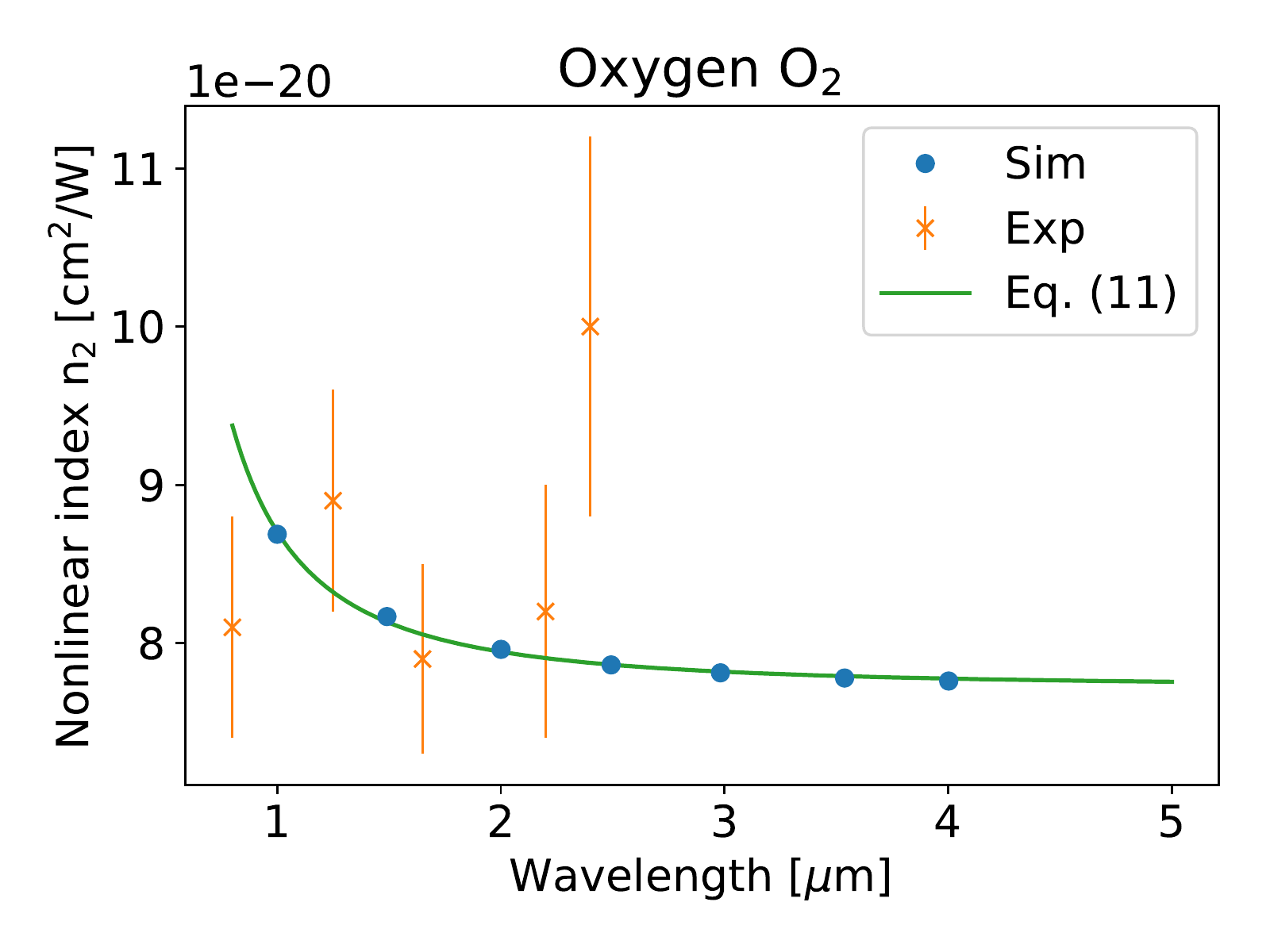}
  \caption{The calculated values (Sim) for the nonlinear refractive
    index $n_2$ of O$_2$ are compared to measured values
    \cite{zahedpour2015n2} (Exp) and to Eq. \eqref{eq:new} with the
    appropriate parameters.}
  \label{fig:oxygen-n2}
\end{figure}

\begin{figure}[h]
  \centering
  \includegraphics[width=0.8\linewidth]{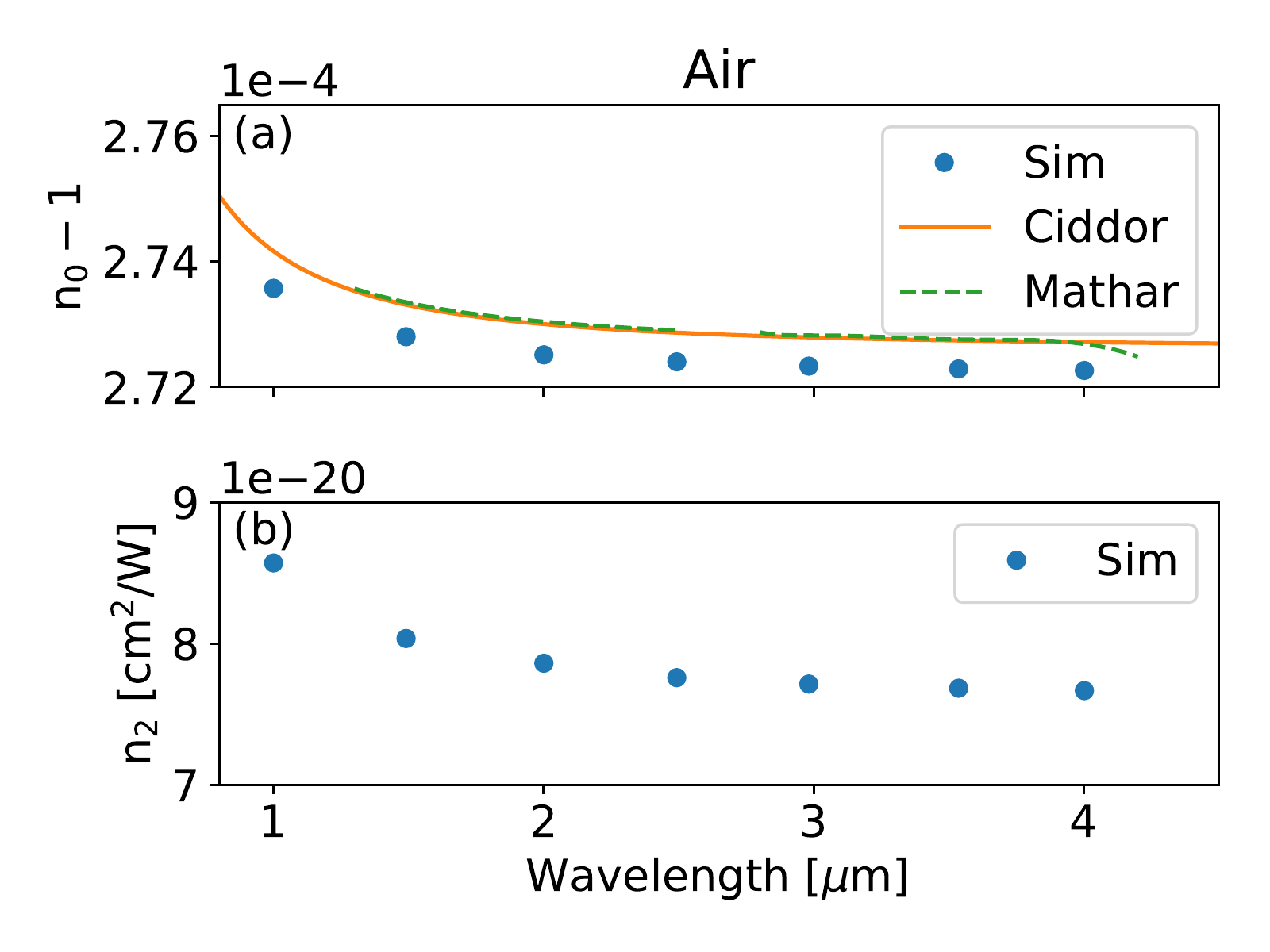}
  \caption{(a) The calculated linear index values for air (Sim) are
    compared to the linear index curves of Ciddor 1996
    \cite{Ciddor:96} and of Mathar 2007 \cite{Mathar-mir-index}. (b)
    The nonlinear index values $n_2$ of air using a proportional
    combination of values from N$_2$ and O$_2$.}
  \label{fig:air-n0-n2}
\end{figure}

A simple model for the optical properties of air can be constructed
using a combination of the calculated susceptibilities for N$_2$ and
O$_2$:
$\chi^{(1)}_{air} = 0.8 \chi^{(1)}_{{\rm N}_2} + 0.2 \chi^{(1)}_{{\rm
    O}_2}$ and
$\chi^{(3)}_{air} = 0.8 \chi^{(3)}_{{\rm N}_2} + 0.2 \chi^{(3)}_{{\rm
    O}_2}$.  In Figure \ref{fig:air-n0-n2}a, we compare the calculated
values of $\chi^{(1)}_{air}$ to index curve Ciddor 1996
\cite{Ciddor:96} (valid from $0.23-1.69$ $\mu$m) and index curves Mathar
2007 \cite{Mathar-mir-index} (valid in ranges $1.3-2.5$ $\mu$m and
$2.8-4.2$ $\mu$m).
We find remarkably good agreement and therefore we can recommend the
use of these curves for modeling the linear properties of air within
the MIR wavelength regime of $1-4$ $\mu$m. In Figure
\ref{fig:air-n0-n2}b, we plot calculated values of the nonlinear index
for air and recommend the use of these values in simulations of the
propagation of MIR laser pulses.

\section{Summary}
\label{sec:summary}
Using ab-initio calculations based on TDDFT, we have calculated the
linear and nonlinear refractive indices for N$_2$, O$_2$, and air for
wavelengths of $1-4$ $\mu$m. Close agreement between the experimental
and calculated values of the linear index demonstrates that it is
possible to extend the commonly-used linear index curves into the MIR
region without modification. We also found that the calculated
nonlinear index values $n_2$ for N$_2$ and O$_2$ are in good agreement
with experimental values up to 2.4 $\mu$m, and our calculations
provide new values for wavelengths up to 4 $\mu$m. We showed that the
predictive formulas for the nonlinear index using Miller's rule tend
to underestimate the dispersion of $n_2$ at long wavelengths, and we
proposed an empirical, Sellmeier-type, fit instead.

Our results show that a fully time-dependent calculation of the
molecular response to a strong field can be used to reliably extract
linear and nonlinear susceptibilities for intensities up to
$5\times 10^{13}$ W/cm$^2$, as long as we correct for higher-order
contributions to the dipole spectrum at a given frequency. Above
$5\times 10^{13}$ W/cm$^2$ the ionization-induced depletion of the
ground state starts to influence the calculation and the extracted
susceptibilities are no longer intensity-independent. Our results
provide a benchmark for future experimental and theoretical
determination of the linear and nonlinear refractive indices in the
MIR spectral range, and will allow for accurate calculations of
phenomena involving long-distance propagation in air.

\textbf{Acknowledgement:} We acknowledge fruitful discussions with
Paul Abanador, Ken Lopata and Ken Schafer. This material is based upon
work supported by the Air Force Office of Scientific Research under
MURI Award No. FA9550-16-1-0013.

\bibliography{octopus}

\end{document}